\documentstyle[12pt]{article}

\begin{document}
\begin{center}{\Large{\bf Ambiguities in statistical
calculations of nuclear fragmentation}}\\
\vskip 1.0cm
S. K. Samaddar$^1$, J. N. De$^{1,2}$ and A. Bonasera$^{3}$\\ 
$^1$Saha Institute of Nuclear Physics, 1/AF Bidhannagar, Kolkata 700064, India\\
$^2$The Cyclotron Institute, Texas A$\&$M University, College Station,\\
Texas 77843, USA\\
$^3$Laboratori Nazionali del Sud, INFN, via S. Sofia 44 95125 Catania,Italy\\
\end{center}
\begin{abstract}
The concept of freeze out volume used in many statistical approaches 
for disassembly of hot nuclei leads to ambiguities. The fragmentation pattern
and the momentum distribution (temperature) of the emanated fragments
are determined by the phase space at the freeze-out volume where the
interaction among the fragments is supposedly frozen out. However, 
to get coherence
with the experimental momentum distribution of the charged particles, one
introduces Coulomb acceleration beyond this freeze-out. To be consistent,
we investigate the effect of the attractive nuclear force beyond this
volume and find that the possible recombination of the fragments alters
the physical observables significantly casting doubt on the consistency
of the statistical model.
\end{abstract}
\vskip 1.0cm
PACS Number(s): 25.70.Pq 24.10.Pa
\newpage

Multifragmentation of nuclear systems in energetic nuclear collisions
serves a novel window for understanding the properties of hot nuclear
matter. It has a sensitive bearing on the 
nuclear equation of state (EoS)
\cite{bona,sto}, focusses our attention on the possibility of liquid-gas
phase transition in finite and infinite nuclear systems \cite{poc,bor}
and from the production of rare isotopes helps for a better understanding
of the nucleosynthesis in astrophysical context \cite{ish}.
Various statistical models have been suggested to explain 
the phenomenon of nuclear multifragmentation;
dynamical models have also been proposed which we will not discuss 
further in this work\cite{bona,sto,bon1}. Broadly the statistical 
approaches are classified in two groups, 
namely, sequential binary decay (SBD) \cite{fri,ric} and one-step prompt
multifragmentation (PM) \cite{gro1,bon2}. It is generally believed that at
low excitation energy, fragmentation proceeds through SBD whereas at
relatively higher energies, it is possibly a one-step break-up process.
Nuclear disassembly in the PM picture has been viewed as a statistical
process and different genres of statistical hierarchy have been employed
to explain the physical process, from grand canonical \cite{koo},
canonical \cite{bon2} to micro-canonical \cite{gro1}. In all these
statistical calculations, a freeze-out volume, around $3V_0$ to
$8V_0$ ($V_0$ being the normal nuclear volume of the fragmenting
system) is employed when the PM process takes place. The fragments
so generated are the primary fragments which are in general in
the excited states. Secondary decay from these hot fragments have been
taken into account \cite{kol,hua}.  Furthermore, in the PM models the fragments
are generally distributed in the freeze-out volume and Coulomb trajectories are
calculated for an improved description of the momentum distribution
of the charged fragments. 
At high excitation energies a collective motion further need to be added 
to the fragments \cite{bona,sto}. These are inconsistent with the assumption
of {\it freeze-out} and is a pointer to the inadequacy of the statistical
model. On the same footing, the action of interfragment nuclear force
beyond freeze-out needs to be considered. The evolution of the
fragments under the combined action of the Coulomb and nuclear fields has not
been considered so far except the one reported in ref.\cite{pal1}. 
In the exit channel motion, two fragments, when close enough to be
under the influence of the nuclear force may recombine to produce 
an excited heavier fragment which may or may not decay further.
In \cite{pal1}, these effects were taken into account and it was 
found that the yield of relatively heavier fragments is enhanced
significantly.  Of course this implies that the original yield and momentum 
distributions given
by the statistical model are altered and in order to get, for instance, 
roughly the original yield, one has to adjust the parameters
including the collective flow. The large success of 
statistical models
(without recombination) in reproducing experimental data, has been 
tacitly assumed as a
proof of equilibration in nuclear collisions.  Including the nuclear force 
(which is a must) for
such small freeze out volumes might significantly alter this 
'idola tribus' (or belief) and the
role of dynamics, so far hidden under the carpet of a few parameters 
and ideal assumptions,
must be reconsidered.  In this work we will discuss the role of nuclear 
forces in the model beyond the freeze-out;
because we are adding some minimum of dynamical effects, we will 
dub our approach as
Dynamical Statistical Fragmentation Model (DSFM). In a later work we will
discuss the flow effects\cite{pal1,pal2}.  The situation now is somewhat
similar to fission, where one utilizes statistical models to determine the mass
distribution at the saddle point and evolves the system including Coulomb, 
nuclear and 
even friction forces.  At variance with fragmentation, the 
system evolves after the saddle point
 thus the mass distributions given by the statistical models are not 
altered.  Including radial
flow in DSFM will roughly cancel the effect of recombination and give 
a picture similar to 
the fission one. 

Isotopic yields from 
multifragmentation have been employed to infer about
important physical 
observables like the temperature of the fragmenting system
\cite{alb} and the associated liquid-gas 
phase transition in finite nuclei.
The effect of recombination has so far been ignored in drawing these
inferences. The results reported in\cite{pal1} are restricted to a fixed
freeze-out volume $V_f$ and 
excitation energy $E^*$; no attempt was made to
look into the consequences
 of the changed isotopic yield on the physical
observables 
after recombination. In the present communication, the effect of
 recombination with the variation in $V_f$ as well as in $E^*$ has been
addressed in some detail and we find that the importance of 
recombination on
the multifragmentation scenario can not be ignored, 
further adding to the
ambiguities of the statistical approaches discussed above. 

The model
employed in the present calculation is the same as that in\cite{pal1}. For
the sake of completeness, only the salient features 
of the methodology are
discussed here. In the first step, the fragment 
multiplicities $n_i$ for the
various fragments are evaluated in the 
grand canonical model (GCM). They are
given by 
\begin{eqnarray}
n_i = V_f\left(\frac {mA_i}{2\pi\hbar
^2\beta}\right)^{3/2} \phi_i(\beta)exp[-\beta (B-B_i+V_i-\mu_n N_i-\mu_p
Z_i)],
\end{eqnarray}
where $\beta$ is the inverse of the temperature $T$, $m$
the nucleon mass, $A_i, N_i$ and $Z_i$ are the mass, neutron and 
charge
numbers of the fragment species $i$, $B$'s are the ground state binding
 energies of the fragmenting system and the generated 
species, $\mu$'s are the
nucleonic chemical potentials and $\phi_i(\beta)$'s are the internal
partition function. The internal partition function is calculated with the
assumption that the 
excitation of the fragment is below the particle emission
threshold. 
 The single particle potential $V_i$ is the sum of the Coulomb
and 
nuclear interaction of the $i$th fragment with the rest of the
fragments 
and is evaluated in the complementary fragment approximation
\cite{gro2,sat}. Employing the GCM fragment formation probability
$p_i=n_i/\sum n_i$, microcanonical 
events are generated following the method
similar to that 
given by Fai and Randrup \cite{fai}. After generation of
fragments in a microcanonical event, the fragments are placed in a
nonoverlapping manner 
within the freeze-out volume. The {\it microcanonical
temperature} is 
evaluated from energy conservation. The fragment velocities
are generated 
from a Maxwell-Boltzmann distribution commensurate with the
microcanonical temperature. 
At this stage the role of the statistical model
is over, but a further 
Coulomb acceleration is now considered which is in
contrast to the statistical assumption. 
Even if the introduction of dynamics
in the model is accepted, one should be
 consistent and include the nuclear
forces as well since the nuclear surfaces in 
the freeze out volume are rather
close to each other. Evaluation of the Coulomb interaction is
straightforward; the nuclear part of the
 interfragment interaction in the
exit channel is broadly classified
 in three groups depending on the masses of
 the fragments. The details are given 
in ref.\cite{pal1}. Two fragments in
the exit channel are assumed to
 coalesce when they touch each other. If
the 
excitation energy of the coalesced fragment is above the particle 
emission
threshold (taken as 8 $MeV$), the fragment is assumed to 
undergo binary
decay; 
the decay probability is calculated in the 
transition state model of
Swiatecki \cite{swi}. 

To study the effect of recombination in nuclear
multifragmentation 
we have considered $^{197}Au$ as a representative system.
In order 
to see the effect of recombination on excitation energy,  the
calculations 
have been performed at $E^*/A$=3, 4 and 5 $MeV$ with a fixed
freeze-out 
volume $V_f=6V_0$. Volume effects have also been considered
with $E^*/A$ fixed at 4 $MeV$. For generation of microcanonical
ensemble, 
typically $10^5$ events have been used. Since we have assumed
that 
the fragments are produced in the particle stable states, the charge 
or
mass distribution is decided at the very onset of fragmentation 
if there is
no recombination. The recombined complex may have excitation 
above the
particle emission threshold and they may 
undergo sequential binary decay in
flight till a particle-stable state is reached. 
In the panels (a), (b) and
(c) of Fig.1, the charge distributions 
at different excitation energies at
$V_f = 6V_0$ are displayed. Except 
for the very light charge particles, the
fragment yield is substantially enhanced. 
At the lowest excitation energy
considered (3 $MeV/A$), the yield of 
very heavy fragments is found to be
somewhat reduced. The neutron yield is enhanced at all the 
excitation
energies considered. This behavior results from a 
delicate interplay between
fragment recombination and subsequent 
binary decay. It is expected that with
reduction in freeze-out volume, the 
recombination effect would be more
 prominent. This is apparent from 
Fig.1(b) and 1(d). This is further evident
from the left panel of 
Fig.2 where the charge distribution has been displayed
 for a very large freeze-out volume (16$V_0$)
 at the same excitation energy of
4 $MeV$ per particle. One would expect the recombination effect 
to be minimal
at this large freeze-out volume, however, we find that though it is
significantly reduced, it is not negligible, particularly 
for fragments with
$Z > 10$. In order to understand the persistence 
of the recombination effect
at this large freeze-out volume, we 
have calculated the surface to surface
separation ($S$) of the different 
fragment pairs ($N_{pair}$) produced in a
disassembly event. In the 
right panel of Fig.2, the average number of
fragment pairs ($\langle N_{pair} \rangle$) 
present within the separation
distance $S$ is displayed for different
 freeze-out volumes at $E^*/A$ = 4
$MeV$.  The fragment pairs within the 
nuclear force range (taken as 1.4
 $fm$  shown as the horizontal dotted line)
 are potential candidates to
 undergo recombination. 
It is seen from the figure that even at $V_f =
16V_0$, there are 
significant number of fragment pairs within the nuclear
force field.

The knowledge of temperature of the disassembling system is crucial in
drawing many important physical inferences such as liquid-gas 
phase
transition. There is no direct way to measure the temperature in such
processes; a number of thermometers have been proposed to that end.
Experimentally, it has been the usual practice to resort to the
 isotopic
double-ratio \cite{alb} to extract the 
temperature which is based on the
statistical multifragmentation model 
with certain approximations. If the
 isotopic yield changes due to recombination,
 the extracted temperature based
 on this model is bound to be erroneous.
 To investigate this aspect, we have
calculated temperatures from different 
isotopic double-ratios at a number
of excitation energies with and without 
the effects of recombination. This is
displayed in panels (a), (b) 
and (c) of Fig.3. It is found that
 the
temperatures extracted without recombinaton are consistent with 
the
excitation energies, however, with inclusion of recombination 
effects, the
extracted temperatures from the isotopic double-ratios decrease
dramatically. Recombination introduces a multitude of 
low temperature sources
in the system 
which may be responsible for the reduction in the temperature
observed. An anomalous fall 
in temperature at $E^*/A =$ 4 $MeV$ is also seen
for all the double-ratio thermometers. 
The temperature extracted after
recombination are, however, found to be 
not too sensitive to the excitation
energy(3-6 $MeV$ per nucleon) that we have considered.
  The dependence of
the double-ratio temperature 
on the freeze-out volume is displayed in
Fig.3(d). 
We have chosen a representative thermometer $(d/t)/(^3He/^4He)$ at
an excitation energy
 $E^*/A$ = 4 $MeV$. Even at the very large freeze-out
volume of 16$V_0$, the temperatures extracted without and with
recombination 
effects are appreciably different, but a very slow approach to
a common 
temperature with increasing $V_f$ is apparent from the figure.

 To sum up, the effect of recombination of fragments on the
charge 
distributions and isotopic double-ratio temperatures in a
nuclear 
disassembly process in the statistical model has been investigated
in  this paper 
at different excitation energies and freeze-out volumes. The
effect is found to 
be very significant for both the observables. With
recombination, the yields for relatively heavy
 fragments are appreciably
enhanced at all the 
excitation energies we have considered. This persistence
of larger yield continues even at a freeze-out volume 
as large as $16V_0$.
With recombination considered, the isotopic 
double-ratio temperature is
reduced dramatically. 
The extracted temperatures without recombination are
found to be not too different 
from those obtained in the Fermi gas model,
however, with inclusion 
of recombination, in the excitation energy range of
3-6 $MeV$ per 
nucleon that we have investigated, the temperatures are found
to be $\sim 4$ $MeV$ and not too sensitive
 to the excitation energy. At
relatively higher excitations, 
the collective flow and an improved dynamics
in the 
fusion process are 
likely to play a very important role and should 
be
taken into account. 
 We, however, stress that the statistical approaches in
a freeze-out scenario should not need any dynamics.
 The large effects seen
with the introduction of dynamics
 (though has been done in an {\it ad-hoc}
manner) even in a large freeze-out 
volume is counter-intuitive and cast
doubt on the applicability of the 
freeze-out concept in the fragmentation
process; 
nonequilibrium dynamical 
features \cite{bona,bon1} should possibly
be incorporated at the very outset.
\vskip 0.7cm
\begin{center}
{\bf Acknowledgement}
\end{center}
S.K.S.  acknowledges the Council of Scientific and Industrial Research
of the Government of India, for the financial support. J.N.D.
gratefully acknowledges the kind hospitality at the Cyclotron Institute
at Texas A $\&$ M University where the work was partially done.

Ths DSFM-code discussed here is 
available and may be requested to one of the
authors. 

\newpage
\centerline 
{\bf Figure Captions}
\begin{itemize}
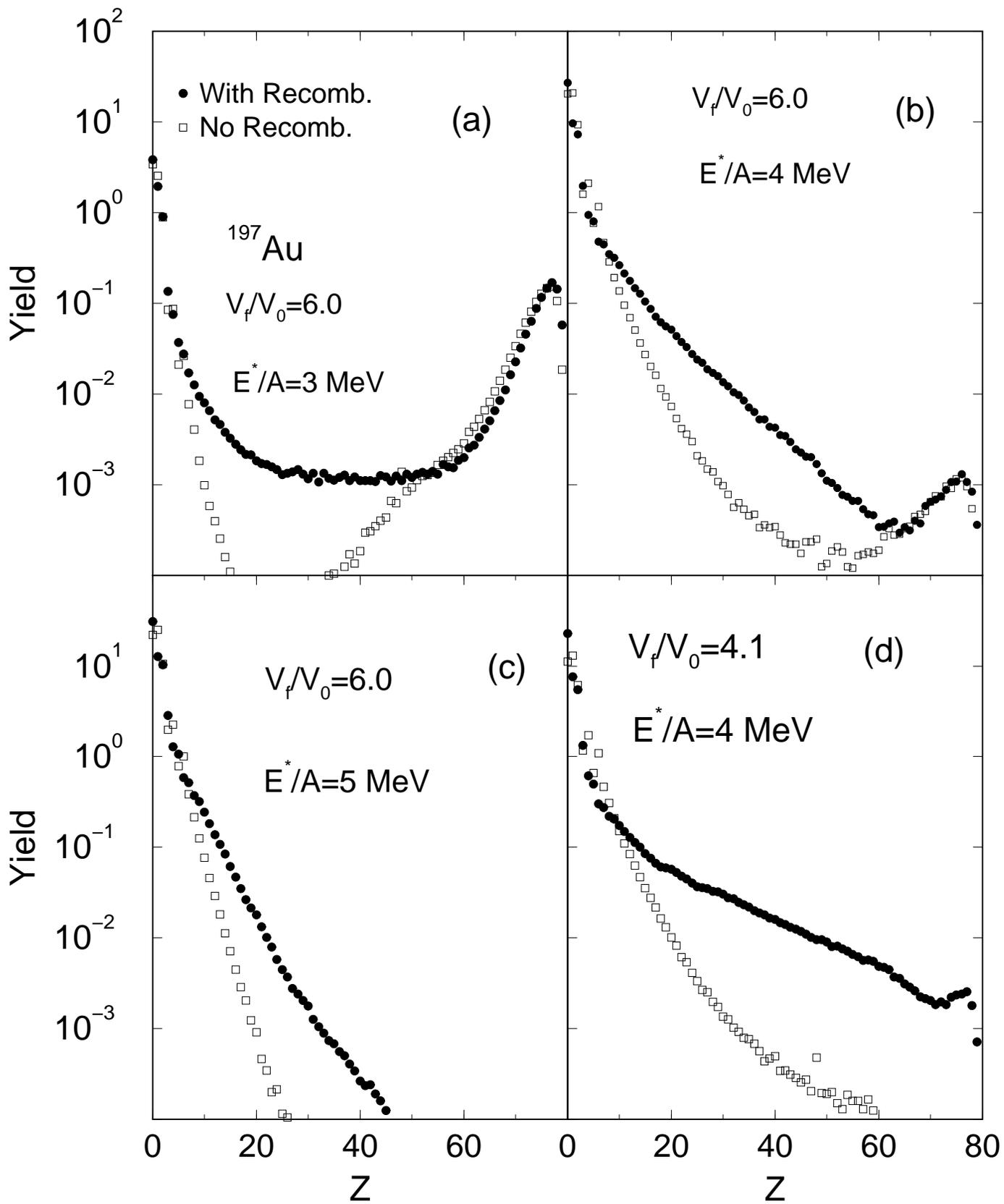
\item[Fig.\ 1] Charge distributions from the
fragmenting system$^{197}Au$ with and without
 recombination at different
excitation energies and freeze-out
 volumes as indicated in the figure.
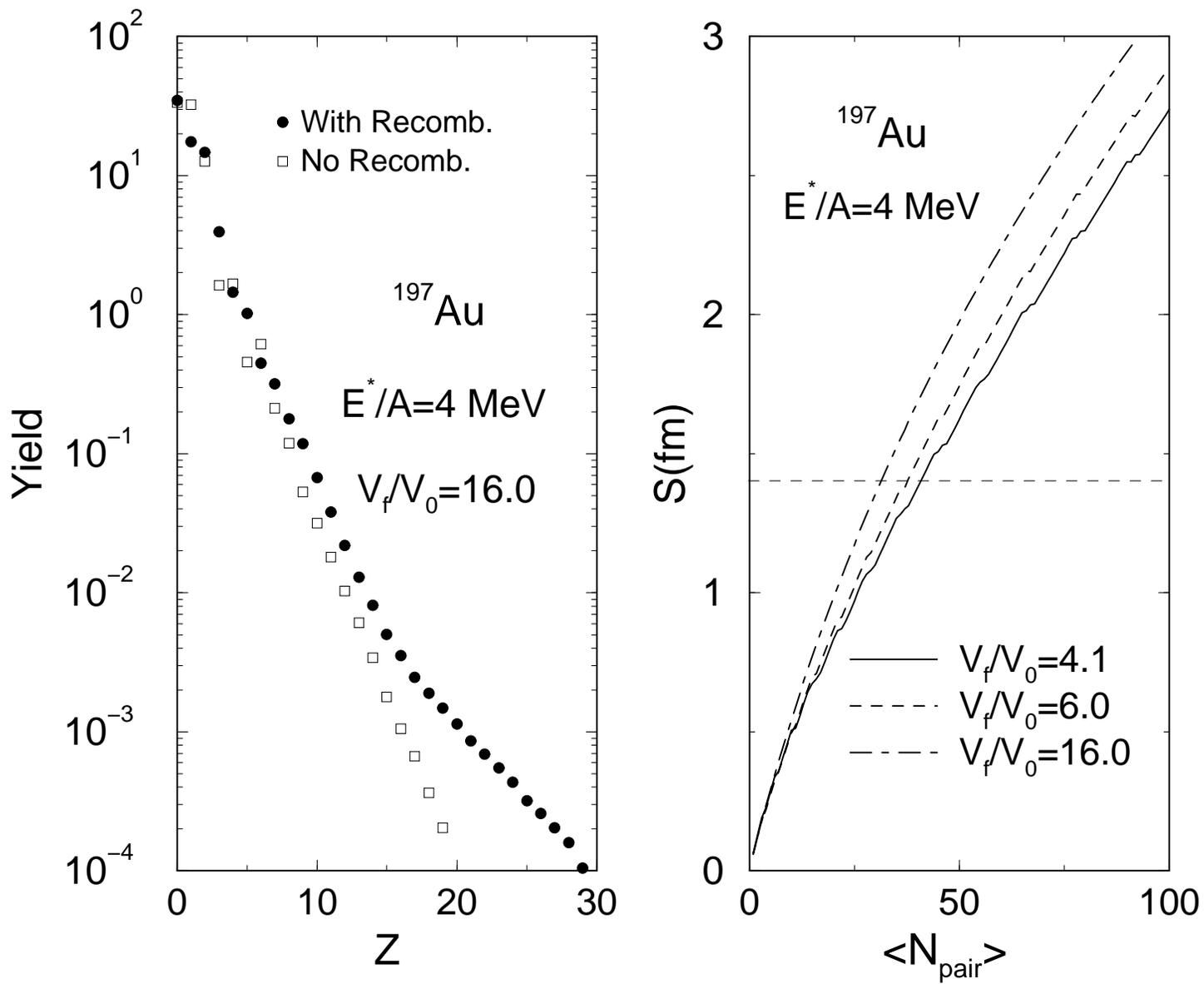
\item[Fig.\ 2] In the left panel the charge distribution
from$^{197}Au$ at an 
excitation energy of 4 $MeV$ per nucleon and$V_f =
16V_0$ is displayed. 
The right panel shows the average number of fragment
pairs within a separation 
distance $S$ at different freeze-out volumes at
the same excitation energy.
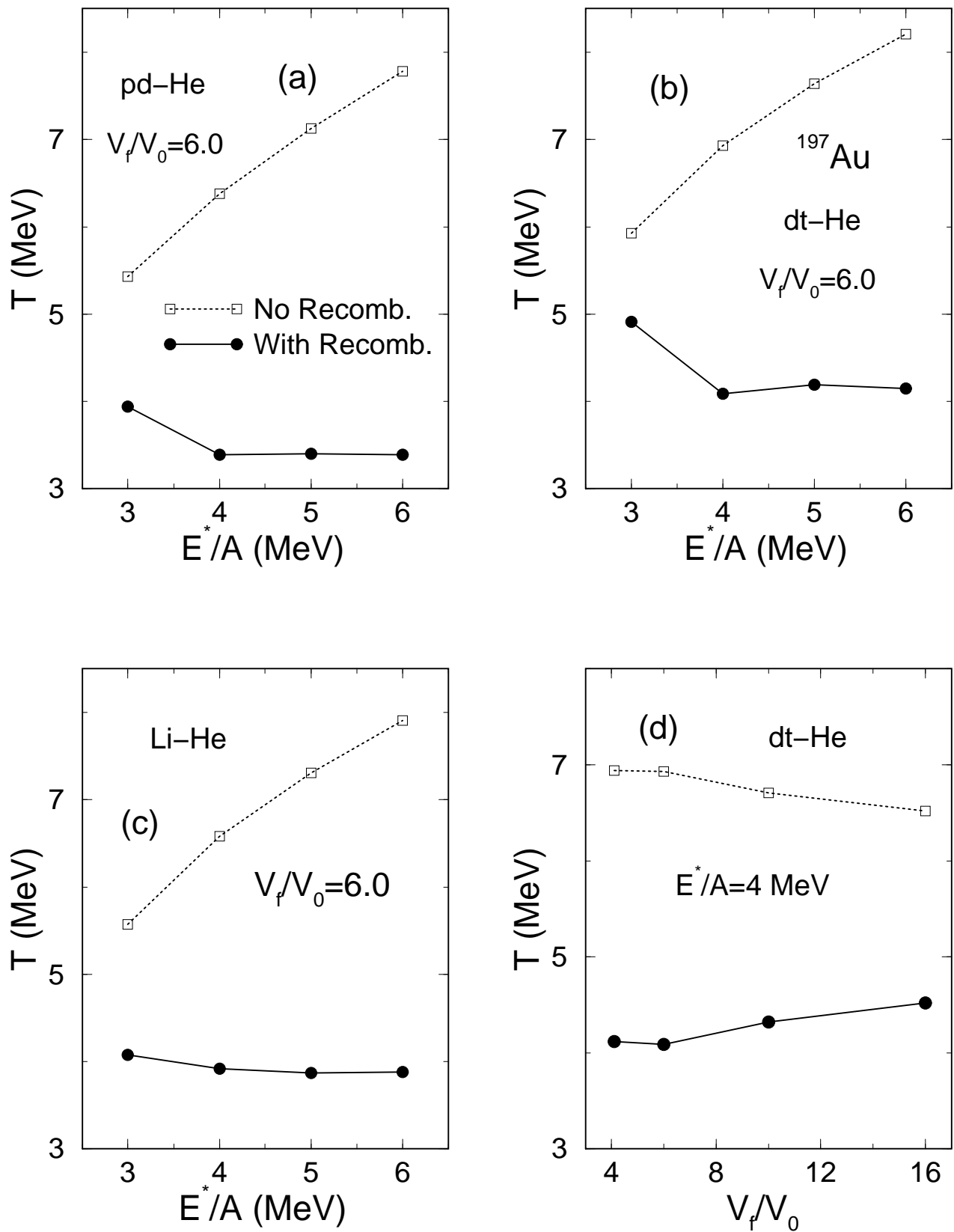
\item[Fig.\ 3] In panels (a), (b) and (c),
different isotopic double-ratio temperatures are shown at various
excitation energies at
 $V_f = 6V_0$ with and without recombination. In panel
(d), the volume dependence of the double-ratio temperature
$(d/t)/(^3He/^4He)$ at $E^*/A$ =4 $MeV$ is
displayed.
\end{itemize}
\end{document}